\documentclass[12pt]{iopart}
\usepackage{graphicx}% Include figure files
\usepackage{dcolumn}% Align table columns on decimal point
\usepackage{bm}% bold math
\begin{document}

\title{Potts Model On Random Trees}% Force line breaks with \\
\author{G.C.M.A. Ehrhardt 
\footnote[3]{To whom correspondence should be addressed (gehrhard@ictp.trieste.it)}
and M. Marsili}
%^  \email{gehrhard@ictp.trieste.it}
%^ \email{marsili@ictp.trieste.it}
%^\affiliation{The Abdus Salam International Centre for Theoretical Physics, Strada Costiera 11, 34100 Trieste, Italy}
\address{The Abdus Salam International Centre for Theoretical Physics, Strada Costiera 11, 34100 Trieste, Italy}
\date{\today}
\begin{abstract}
We study the Potts model on locally tree-like random graphs of arbitrary degree distribution.
Using a population dynamics algorithm we numerically solve the problem exactly.
We confirm our results with simulations. 
Comparisons with a previous approach %\cite{gdmisingonrgs,dgmpottsonrg} 
are made, showing where its assumption of uniform local fields breaks down for networks with nodes of low degree.
\end{abstract}
\maketitle

Recent years have seen intensive interest in the statistical physics community in complex networks.   Random graphs with a given degree distribution (often also called the configuration model) \cite{newmanonrandomgraphs,rgswitharbitrarydegreedistributionsandtheirapplications}, small world graphs \cite{ballssmallworld,wattsandstrogatzsmallworld,wattssmallworldsbook} and various scale-free graphs based on preferential attachment \cite{priceoriginalprefatt,albertandbarabasiscience} have been extensively studied \cite{dorogovtsevandmendesreview,newmanreview,albertandbarabsaireview}.  
Many researchers have also considered processes taking place on complex networks, such as percolation \cite{NWsmallworldscalingandpercolation,percolationonRG1,percolationonRG2}, disease spreading \cite{ballssmallworld,diseaseoninternet_nothresholds}, self-organised critical models \cite{baksneppenonsmallworld,BTWsandpileonsmallworld,gbianconi_cloggingofnets}, and various games \cite{games1,games2}.  These processes have in general been studied in the past on regular lattices, the current interest being in how the topology of the complex network affects the processes.
The Ising model and its generalisation the $q$-state Potts model have also recently been studied on random graphs of a given degree distribution \cite{gdmisingonrgs,dgmpottsonrg,leoneetal}.  These models are standard models of statistical physics, displaying continuous (for Ising) and first order ($q>2$) phase transitions \cite{baxterpaper} in mean field.  Their behaviour on complex networks, as well as being of interest in themselves, also gives clues as to how more complicated processes will behave.

The solution of the Potts model on a Bethe Lattice has been known for some time \cite{baxtersbook}.  More recently, Dorogovtsev, Goltsev, and Mendes have extended the solution to random trees \cite{dgmpottsonrg}.  This includes the configuration model for large network sizes since in these cases the graph is locally tree-like.  Their approach, which is tailored to scale-free random graphs, relies on an effective field approximation which accounts for the effect of the nodes with the highest degree. This is indeed able to capture the correct nature of the phase transition and provides a qualitatively correct behaviour.

In this paper we describe a method of numerically solving for the local fields without making any approximation. This shows that, at odds with the effective field approximation \cite{dgmpottsonrg}, local fields and the magnetization indeed depend on the local geometry and hence fluctuate with it. The effects show their relevance, for example, in a recent study where the Potts model on a random graph was used to model coordination games on an evolving network \cite{gmf_inprep}. 

We compare our results with the method of \cite{dgmpottsonrg} and with simulations.  We find that, as expected, the approximation of \cite{dgmpottsonrg} gives poor results for graphs containing nodes of low connectivity but works well when all nodes have large connectivity.

\section{Potts Solution}
The $q$-state Potts model has the Hamiltonian
\begin{equation}
{\cal H}=-J \sum_{<i,j>} \delta_{\sigma_i,\sigma_j} -{\rm b} \sum_{i} \delta_{\sigma_i,1}
\end{equation}
where the $\sigma$s take one of the integer values from $1$ to $q$ and the sum $\left<i,j\right>$ is over neighbouring nodes.  Here the applied magnetic field $b$ acts on the (arbitrarily chosen) state $1$.
The partition function is then:
\begin{equation}
Z= \sum_{\{{\bf \sigma}\}} \exp \left( K \sum_{<i,j>} \delta_{\sigma_i,\sigma_j}  +H \sum_{i} \delta_{\sigma_i,1} \right)
\end{equation}
where $K=\beta J$ and $H= \beta {\rm b}$, $\beta = 1/(k_B T)$, $k_B$, $T$ being Boltzmann's constant and temperature respectively.
The local magnetisation $m_i$ is given by
\begin{equation}
m_i=\frac{ q \left< \delta_{\sigma_i,1} \right> -1    }{   q-1  }
\end{equation}
where the angled brackets denote the Gibbs average.  The total magnetisation is then the average of $m_i$ over all nodes. 

We are interested in the Potts model on a (connected) random graph of $n$ nodes with specified degree distribution $P(k)$. These are locally tree-like graphs in the sense that the sub-network of all nodes at a distance less than $\ell$ steps from any particular root node is very similar to a tree, for small $\ell$. In particular, this is true for $\ell\ll \bar d$, where $\bar d$ is the diameter of the network, which is known \cite{sizeofloopsgoesaslnn} to behave as $\bar d\sim\log n$ if the second moment of $P(k)$ is finite and $\bar d\sim\log\log n$ if the second moment diverges, but the first is finite. The strategy of the solution relies on the following idea: We take a particular site as the root and show that its statistical properties can be derived from those of their neighbors. In turn, the properties
of the neighbors are defined by those of second neighbors and so on. As long as the graph is tree-like, i.e. for distances $r\ll \bar d$ from the root, the statistical properties of the sites involved in the $r^{\rm th}$ iteration -- those at distance $r$ from the root -- are independent, and are captured by a simple transfer matrix. If this transfer matrix has a fixed point, the statistical properties of bulk nodes will be described by the fixed point irrespective of the boundary conditions which are set at the leaves of the tree (i.e. at distance $r\sim \log n$). This will be true, in particular, if the boundary conditions are drawn from the fixed point distribution which describes the statistics of bulk sites. Notice that this argument requires the diameter over which the network is tree-like be much larger than the correlation length over which the transfer matrix converges. Hence for finite graphs we expect the local tree approximation to perform poorly for graphs with $\bar d\sim \log\log n$ and also close to the critical point where the convergence of the iterative process is slowed down.

For a tree, following the notations of Ref. \cite{dgmpottsonrg}, we can solve $Z$ iteratively, we define 
\begin{eqnarray}
g_{1,i}(\sigma_0)= \sum_{\{\sigma_l\}} e^{\sum_{<l,m>} K \delta_{\sigma_l,\sigma_m} %\nonumber \\
+ K \delta_{\sigma_0,\sigma_i} +H \sum_{l} \delta_{\sigma_l,1}}
\end{eqnarray}
Where the sum over $l$ and $m$ is {\it only} over nodes in the sub-tree whose root node is $i$, including $i$.
Thus
\begin{equation}
Z=\sum_{\sigma_0} e^{H \delta_{\sigma_0,1}} \prod_{i=1}^{k_0} g_{1,i}(\sigma_0)
\end{equation}
and
\begin{equation}
\label{m0eqn}
\left<m_0\right> = \frac{ \exp(H) -\prod_{i=1}^{k_0} x_{1,i}  }{ \exp(H) +(q-1) \prod_{i=1}^{k_0} x_{1,i} }
\end{equation}
where $x_{1,i}=g_{1,i}(\alpha)/g_{1,i}(1)$. Likewise, we can define the quantity $g_{r,i}(\sigma_{r-1})$ for any node $i$ at a distance $r$ from the root, as the partial sum of the partition function on the sub-tree ensuing from that node, where $\sigma_{r-1}$ is the value of the spin at distance $r-1$ where the sub-tree is connected. 
This satisfies the recursion relation
\begin{equation}\label{gni}
g_{r,i}(\sigma_{r-1})=\sum_{\sigma_r} e^{K \delta_{\sigma_r,\sigma_{r-1}}+H\delta_{\sigma_r,1}}
\prod_{j=1}^{k_{r,i}-1} g_{r+1,j}(\sigma_r)
\end{equation}
where $k_{r,i}$ is the degree of node $(r,i)$. Notice that $g_{r,i}$ depends on the value of the spin $\sigma_{r-1}$ at the previous level. 
Writing down these equations for $\sigma_{r-1}=1$ and $\sigma_{r-1}=\alpha>1$ and taking the ratio
one finds an iterative equation for 

\begin{equation}\label{leaf}
  x_{r,i}=\frac{g_{r,i}(\alpha)}{g_{r,i}(1)},~~~~~~~~~\alpha>1
\end{equation}
which reads
\begin{equation}
\label{fk}
  x_{r,i}=\frac{e^{H}+(q-2+e^K)\prod_{j=1}^{k_{r,i}-1} x_{r+1,j}}
  {e^{H+K}+(q-1)\prod_{j=1}^{k_{r,i}-1} x_{r+1,j}} .
\end{equation}
In principle $x_{r,i}$ should depend on $\alpha$ in Eq. (\ref{leaf}). However it is easy to see that $x_{r,i}$ is independent of $\alpha$ if the node $(r,i)$ is a leaf of the tree. 
Because all $x_{r,i}$ can be computed recursively with Eq. (\ref{fk}) from the leaves back to the root then $x_{r,i}$
is independent of $\alpha>1$ for all nodes. 

From here, making the change of variables $h_{r,j}=-\ln(x_{r,j})$, Dorogovtsev et. al. \cite{dgmpottsonrg} noted that the only dependence on $h$ in the right hand sides of equations (\ref{m0eqn},\ref{fk}) was from a sum over the independent $h_i$s.  They thus made the approximation
\begin{equation}
\label{dmapprox}
\sum_{j=1}^k h_{r+1,j} \approx k \left<h_{r+1}\right> +O(\sqrt{k})  .
\end{equation}

It is possible instead to solve for the distribution of $h$ values in the steady state.  Let $\rho_{r}(h|k)$ be the distribution density of $h_{r,i}$ with $k_{r,i}=k$. This distribution satisfies
\begin{equation}\label{rhohk}
\rho_r(h|k)=\int_{-\infty}^\infty \prod_{j=1}^{k-1} dh_j \tilde\rho_{r+1}(h_j)\delta\left[h-Y\left(\sum_{j=1}^{k-1} h_j\right)\right]
\end{equation}
where
\begin{equation}
Y({\it s})=\ln\left[\frac{e^{H+K}+(q-1)e^{-{\it s}}}{e^{ H}+(q-2+e^K)e^{-{\it s}}} \right]
\end{equation}
and
\begin{equation}
\tilde\rho_r(h)=\sum_{k=1}^\infty \tilde P(k)\rho_r(h|k)
\end{equation}
is the distribution of $h$ on the neighbours of a node, $\tilde P(k) = k P(k)/\left<k\right>$ being the distribution of degrees of neighbours of a node.  Averaging over $\tilde P(k)$ we find a recursion relation for the distribution of fields at the neighbour of a node, $\tilde\rho_r(h)$
\begin{equation}\label{trho}
\tilde\rho_r(h)=\sum_{k=1}^\infty \tilde P(k)\int_{-\infty}^\infty \prod_{j=1}^{k-1} dh_j \tilde\rho_{r+1}(h_j)\delta\left[h- Y\left(\sum_{j=1}^{k-1} h_j\right)\right]
\end{equation}
Well inside the bulk of the tree, we expect that $\tilde\rho_r\to\tilde\rho$ independent of $r$. Once $\tilde\rho(h)$ is found, the distribution of local fields on any site with degree $k$ is given by Eq. (\ref{rhohk}).

\section{Algorithm for Solution}

A simple way to solve equation (\ref{trho}) numerically is the following: Start from a population $h_{i}$, $i=1,\ldots,M$ of $M\gg 1$ values of $h$.  Evolve the population by iteration with the following procedure:
\begin{itemize}
  \item Draw at random $k$ from the distribution $\tilde P(k)$
  \item Draw $k-1$ values of $h$ at random from the population $\{h_{i}\}$ and sum them to get $h_{sum}$. 
  \item replace a random member of the population by
\begin{equation}
  h_{new}=Y\left(h_{sum}\right)
\end{equation}
\end{itemize}
Iterate until convergence.
The magnetisation is then found by averaging the local magnetisations on many nodes,
\begin{itemize}
  \item draw $k$ from $P(k)$
  \item draw $k$ values of $h$ from $\tilde\rho(h)$ and sum them to get $h_{sum}$ and insert this into\begin{equation}
\left<m_0\right> = \frac{ \exp(H) -\exp( -h_{sum} )  }{ \exp(H) +(q-1) \exp( -h_{sum} ) }
\end{equation}
\end{itemize}
Iterate many times and average the results to get $M(T)$.

By contrast, the approximation of Dorogovtsev et. al. \cite{dgmpottsonrg} gives:
\begin{eqnarray}
\label{dmheqn}
\left<h\right> &=& \sum_{k=1}^\infty \tilde P(k)Y( (k-1)\left<h\right> ) \\
M(T) &=& \sum_{k=1}^\infty P(k) \frac{ \exp(H) -\exp( -k \left<h\right> )  }{ \exp(H) +(q-1) \exp( -k \left<h\right> ) } .
\end{eqnarray}
To find the general $M(T)$ these two equations are also solved numerically, albeit for only one value of $h$ rather than a whole population.

\section{Results and comparison with Simulations}
In the following we compare the results of our method with that of Dorogovtsev et. al. and with simulations.  In particular, for zero applied field ($H=0$), we study the $q=10$ state Potts model on random graphs with degree distributions $P(k)$ that are bimodal or truncated power law,
\begin{eqnarray}
\label{pofkbim}
P(k)&=&\frac{\delta_{k,2}+\delta_{k,10}}{2} \\
P(k)&=& A k^{-\gamma} \,\,\,\,\,\,\,\,\,\, {\rm for} \,\,\, k_{min} \le k \le k_{max}    
\label{pofkpow} 
\end{eqnarray}
where $A$ is the normalisation.  

The random graphs for the simulations are produced in the following way:  $n$ nodes $i$ are given degrees $k_i$ drawn from $P(k)$, if $\sum k_i$ is odd then a randomly chosen node has its degree increased by $1$.  Nodes of degree $k$ are imagined to have $k$ unconnected ends.  Next, pairs of ends are chosen at random and linked iff they belong to different nodes which are not already linked.  This is repeated until no unconnected ends remain, in the unlikely event that there remain unconnected ends at nodes which are already joined, the degree of these nodes is reduced.  The graph is then `shuffled' by picking two links, $g-h$, $i-j$ at random and changing it to either $g-i$, $h-j$ or $g-j$, $h-i$ with equal probability.  This is repeated until on average each link has been rewired $\sim 10$ times.  This shuffling is intended to remove any bias due to not allowing multiple linkings \cite{ontheuniformgenerationofrgsincnewman}.  The $P(k)$s are chosen so that there exists a large connected component (giant component) containing almost all nodes, nodes not in the giant component are discarded when calculating the magnetisation since the solution methods described above are for the giant component only.  Network creation took insignificant time in comparison to simulation of the Potts model.
The Potts model is simulated using the Swendsen-Wang algorithm \cite{swendsenwang}.  System sizes of $n=20,000$ $50,000$ were easily simulated and were sufficient to allow comparison with the solutions described above.

\begin{figure}
\includegraphics[width=\columnwidth,clip]{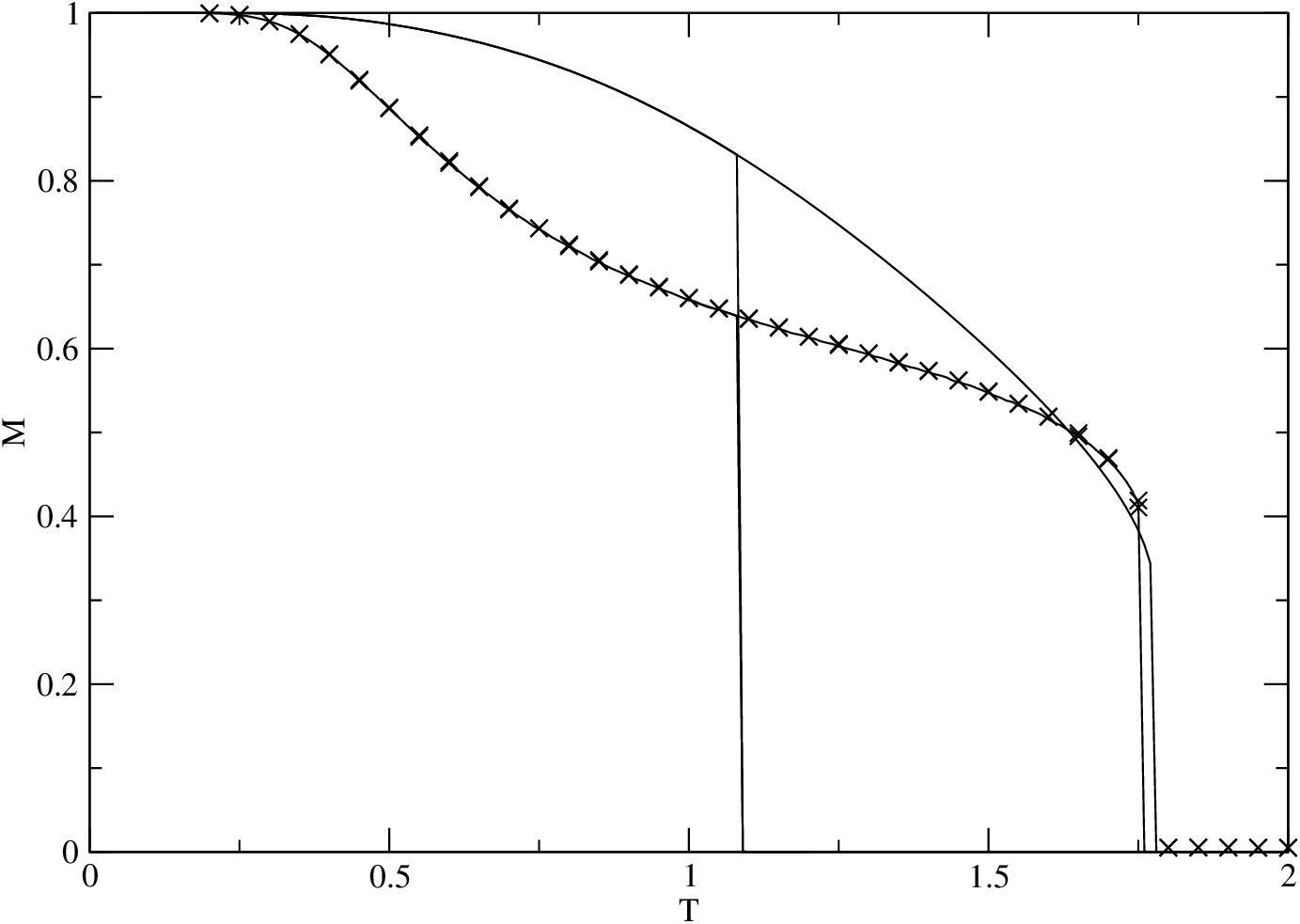}
\caption{ Plot of $M(T)$ for the $q=10$ Potts model on a random graph with bimodal degree distribution (see equation (\ref{pofkbim})) with $J/k_B=1$.  Crosses are simulations for $2$ realisations of a network with $n=20,000$ nodes, starting from an initially fully magnetised state.  Lines are the results of the $2$ methods described, the method introduced here agreeing with simulations to high accuracy.  The vertical line at $T \approx 1.1$ is the transition from the metastable unmagnetised state to the magnetised state, both methods giving the same value of $T_{metastable}$.
\label{bimodalmvst} }
\end{figure}

\begin{figure}
\includegraphics[width=\columnwidth,clip]{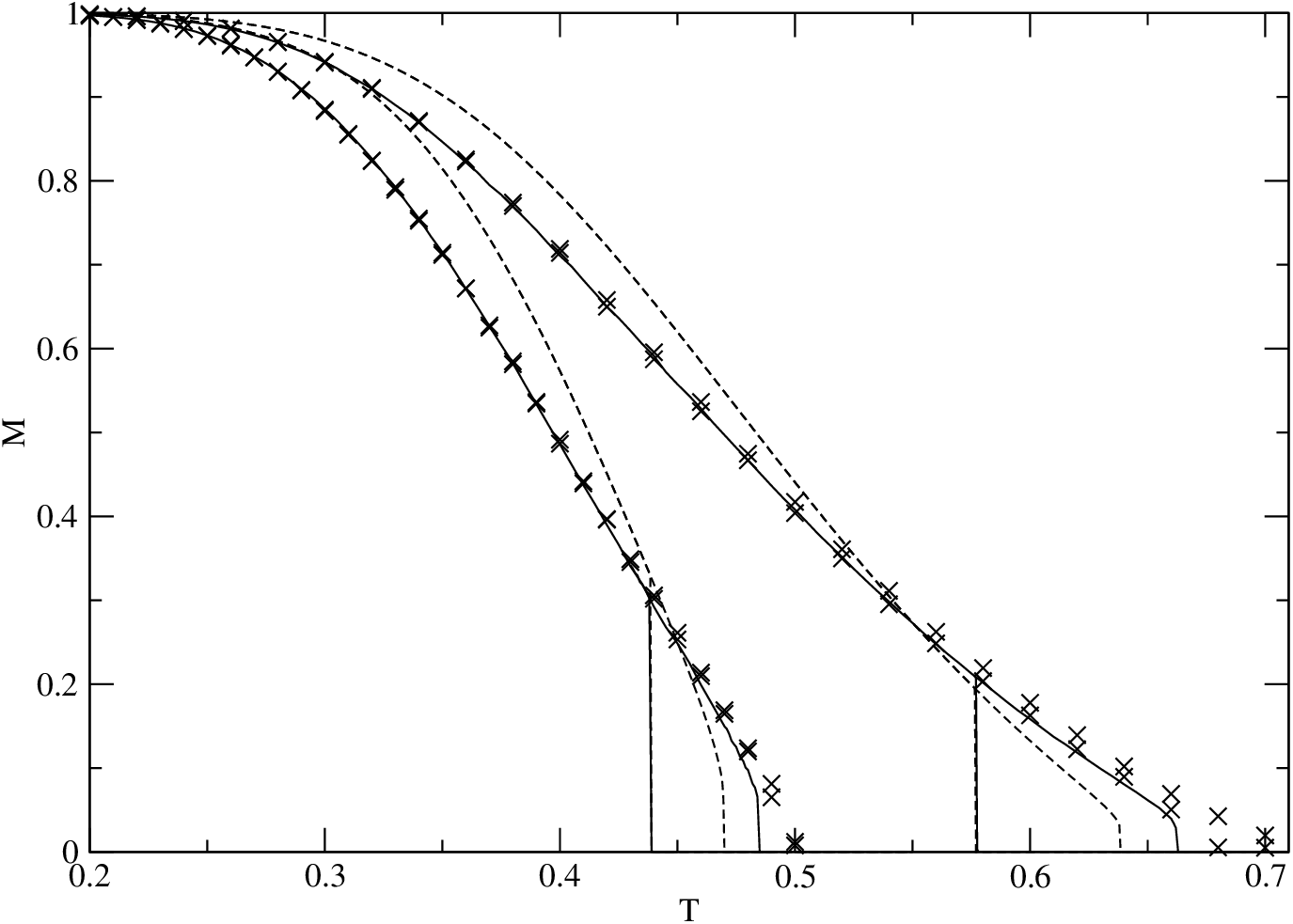} 
%{powerlawalphais4kmaxis101.eps}
\caption{ Plot of $M(T)$ for the $q=10$ Potts model on a random graph with power-law degree distribution (see equation (\ref{pofkpow})) with $J/k_B=1$, $\gamma=3.5$ (rightmost curves) and $\gamma=4$ (leftmost curves), $k_{min}=2$, $k_{max}=100$.  Crosses are simulations for $2$ realisations of a network with $n=50,000$ nodes, starting from an initially fully magnetised state.  Lines are the results of the $2$ methods described, the solid curves are the method introduced here, the dotted lines are the method of Dorogovtsev et. al.  The vertical lines are the transition from the metastable unmagnetised state to the magnetised state, both methods giving the same value of $T_{metastable}$.
\label{powerlawgammais4mvst} }
\end{figure}

In figure \ref{bimodalmvst} we show the results for magnetisation of the bimodal graph as a function of temperature.   This case was chosen specifically because the approximation of Dorogovtsev et. al. would perform poorly due to the small degree ($2$) of half of the nodes and the large separation of the degrees ($2:10$).  As expected, the method of Dorogovtsev et. al. gives an incorrect result for the magnetisation although it's value for the transition temperature $T_c$ is surprisingly close to the correct value.

Figure \ref{powerlawgammais4mvst} shows the results for the power law case with exponents $\gamma=3.5$,$4$ with $k_{min}=2$ and $k_{max}=100$.  In practice the expected number of nodes with $k>60$ is less than $1$ for $n=50,000$
%(see the mathematica file Pofk_forPowerLawCase.nb)
and increasing $k_{max}=100$ does not appreciably change the results of our method. 
Again, the method of Dorogovtsev et. al. differs from the one introduced here and from the simulations.  In particular, the transition temperature is incorrect by $\sim 3\%$.  The simulations do not agree quite as well with our solution as for the bimodal case, particularly near to the transition temperature.  This is expected because, as mentioned earlier, the convergence of the iterative procedure slows down close to the transition.  A signature of this is that even with such large system sizes, the curves obtained for two different realizations of the graph differ slightly. In addition, the agreement improves as $n$ increases up to $50,000$. 

Also shown in figures (\ref{bimodalmvst},\ref{powerlawgammais4mvst}) is the transition from the metastable unmagnetised state at a temperature $T_{metastable}$ less than $T_c$.  $T_{metastable}$ can be found exactly \cite{dgmpottsonrg}
\begin{equation}
T_{metastable} = J/\ln \frac{ \left<k^2\right> +(q-2)\left<k\right> }{ \left<k^2\right> -2\left<k\right> }
\end{equation}
For the continuous transitions, i.e. for $q =2$, there is no metastable region and so $T_{metastable} = T_c$ and thus the method of Dorogovtsev. et. al. gives $T_c$ exactly.

For these results we have used population sizes of $10^5$ and, near $T_c$, $10^6$.  Near $T_c$ and $T_{metastable}$ $\sim 500$ iterations per member of the population were used to ensure convergence, less elsewhere.  Between $T_c$ and $T_{metastable}$ the magnetisation found depends on the initial values of $h_i$, we started from both $h_i \approx 0$ and $h_i \gg 1$, with a slight spread in the distribution of $h_i$ values.  The timescales to produce a full $M(T)$ curve using a standard PC were of the order of minutes for the method of Dorogovtsev et. al. and hours for the method described here.

\begin{figure}
\includegraphics[width=\columnwidth,clip]{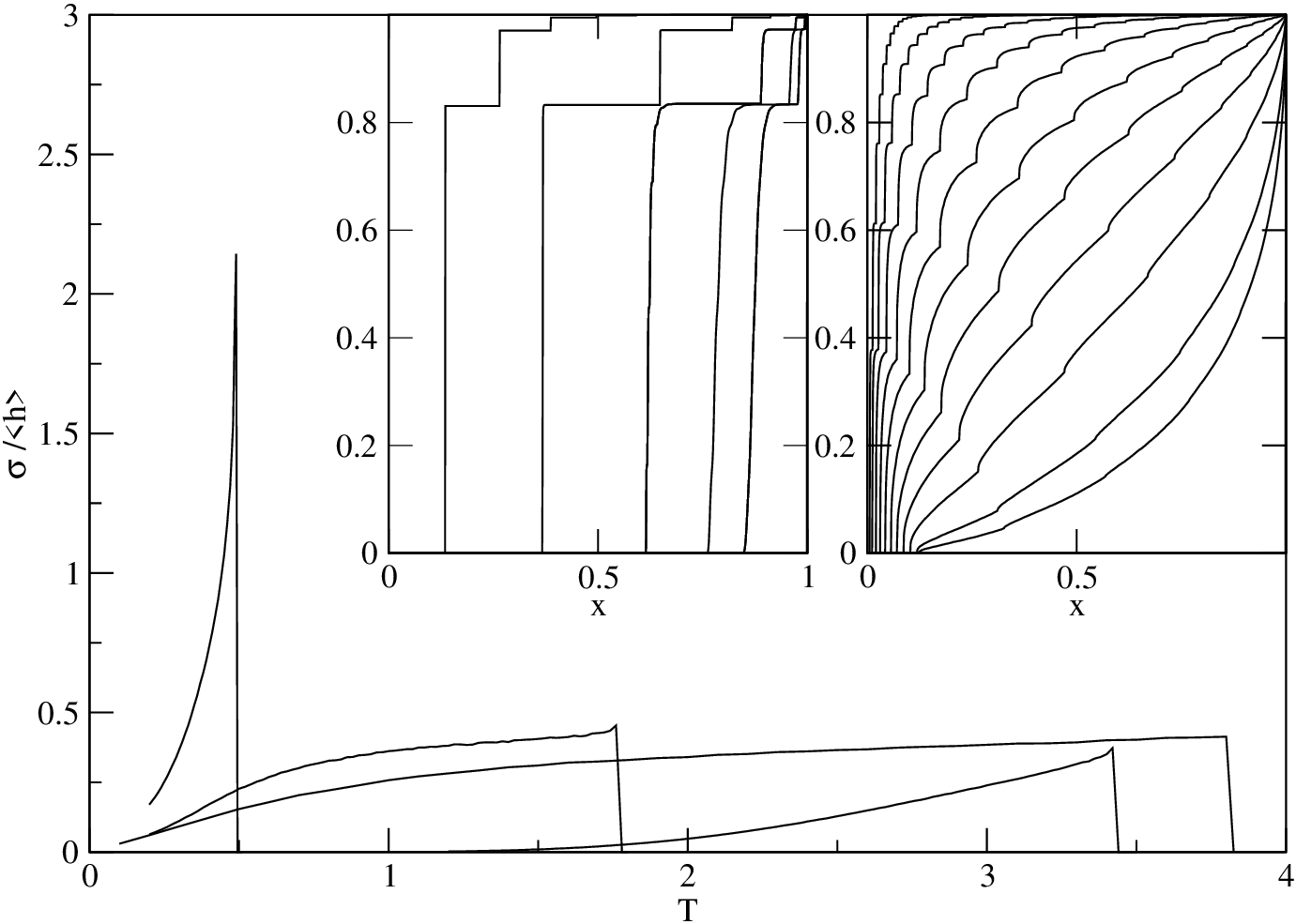}
\caption{ The standard deviation of the effective fields on a neighbouring node, $h_i$, divided by $\left<h\right>$ for $\left<h\right> >0$ as a function of temperature.  Note the drops at the transition temperatures, from left to right, the transitions are for: powerlaw $\gamma=4$ $q=10$ $k_{min}=2$ $k_{max}=100$, bimodal $q=10$ $k=2,10$, bimodal $q=10$ $k=10,20$, and bimodal $q=2$ $k=2,10$.
The insets show the cumulative distribution of effective fields $x$ on a neighbouring node, $\int_0^x \tilde\rho(x') dx'$ for the bimodal random graph with $q=2$ and $k=2,10$ (left inset) and for the powerlaw $\gamma=4$ $q=10$ $k_{min}=2$ $k_{max}=100$ case (right inset).   Various temperatures are shown, $T$ increases from left to right.  The plateau at $y \approx 0.83$ for the bimodal cases is where the $k=2$ nodes are unmagnetised, $y_{plateau}=10/(2+10)$.  
\label{distributionsofhandx} }
\end{figure}

In figures (\ref{distributionsofhandx}) we show the distributions of the effective fields $x_i$ on neighbouring nodes at various temperatures below the phase transition.  It is clear that, except in the fully magnetised or unmagnetised states, there is a wide spread of local field values.  In particular, for the bimodal case, the plateau at $y \approx 0.83$ is because the $k=2$ nodes are unmagnetised whilst the $k=10$ nodes remain magnetised.  This is also consistent with the form of the $M(T)$ curve in figure (\ref{bimodalmvst}).

Figure (\ref{distributionsofhandx}) shows the temperature dependence of the relative spread of local fields, defined as the ratio between the standard deviation $\sigma$ of the effective fields on neighbouring sites $h_i$ divided by $\left<h\right>$ for $\left<h\right> >0$. This is non-zero below the phase transition, and it attains a maximal value at the phase transition. This shows that the approximation of Ref. \cite{dgmpottsonrg} is far from exact. In spite of this, we found that the discrepancy of the results of Ref. \cite{dgmpottsonrg} with the method described here is almost insignificant when the minimal degree $k_{\min}$ is large (say $k_{min}\approx 10$). 
The discrepancy between Dorogovtsev et al.'s approximation and numerical simulations also improves slowly when $q$ decreases, their result being exact for $q=1$ \cite{dgmpottsonrg}. 
The approximation allows some analytic results, e.g. for $T_c$, to be derived within it and it is numerically less demanding and thus significantly faster than ours.

In conclusion, we have presented a method for solving the Potts model on a random graph of given degree distribution.  The method does not make the approximation of Ref. \cite{dgmpottsonrg} and thus is accurate in all cases when the local tree approximation is applicable, giving results in good agreement with our simulations.

%\begin{acknowledgments}
\ack{
Work supported in part by the European Community's Human Potential Programme under contract HPRN-CT-2002-00319, STIPCO.
%\end{acknowledgments}
}

\end{document}